\documentclass[sigconf, nonacm]{acmart}

\author{Niklas Bunzel}
\affiliation{%
  \institution{Fraunhofer SIT / TU Darmstadt / ATHENE}
  \city{Darmstadt}
  \country{Germany}
}
\email{bunzel@sit.fraunhofer.de}

\title{VanillaBench: The Hidden Accuracy Cost of Adversarial Robustness}

\begin{abstract}
Adversarial robustness research has produced hundreds of defended models over the past decade, yet the literature almost universally reports robustness results in isolation: standard (clean) accuracy and adversarial accuracy of the robust model are shown, but the gap to the corresponding vanilla (non-adversarially-trained) model is rarely quantified. We introduce \textbf{VanillaBench}, a systematic benchmark that makes this gap explicit. For every adversarially-trained model catalogued by RobustBench across four threat models (CIFAR-10~$\ell_\infty$~8/255, CIFAR-10~$\ell_2$~0.5, CIFAR-100~$\ell_\infty$~8/255, and ImageNet~$\ell_\infty$~4/255), we compute the accuracy difference $\Delta_{\mathrm{clean}} = \mathrm{acc}_{\mathrm{robust}} - \mathrm{acc}_{\mathrm{vanilla}}$ against multiple vanilla references: the best, median, and top-10 median standard models from Papers with Code, computed over both all entries and no-extra-data entries, the best vanilla model as of the robust model's publication year, and an architecture-matched baseline. Across all 186 robust models, the mean $\Delta_{\mathrm{clean}}$ relative to the best vanilla model ranges from $-7.7$ to $-29.5$ percentage points, and even the single most robust model per track still trails its temporal vanilla counterpart by $4.0$--$21.0$ points. The architecture-matched comparison, which isolates the effect of adversarial training from architectural differences, reveals a mean gap of $-3.5$ to $-17.5$ points. Restricting this architecture-matched comparison to models whose vanilla accuracy is known for the exact same architecture, rather than approximated from a related one, narrows the gap to $-4.0$ to $-14.0$ points. These results demonstrate that the robustness-accuracy trade-off is substantially larger than what is typically conveyed by individual papers. This information is critical for practitioners and decision-makers. When deploying models in real-world settings, the accuracy cost of robustness directly affects business outcomes, yet current publications do not provide the vanilla baseline needed to assess it. We argue that future robustness evaluations should report vanilla-referenced accuracy gaps as a standard component.
\end{abstract}

\keywords{adversarial robustness, accuracy-robustness trade-off, benchmark, image classification}

\usepackage{booktabs}
\usepackage{graphicx}
\usepackage{multirow}
\usepackage{array}
\usepackage{xcolor}

\begin{document}

\maketitle

\section{Introduction}

Since the discovery that deep neural networks are vulnerable to adversarial perturbations~\cite{szegedy2013intriguing,goodfellow2014explaining}, a large and active research community has developed a wide range of defense strategies. One prominent strain of work is \emph{adversarial training}~\cite{madry2017towards, BaiL0WW21, singh2023revisiting, TongJG024, ChhipaVJSSL25}, which hardens models by incorporating adversarial examples directly into the training loop. A complementary strand comprises \emph{preprocessing defenses} that transform or purify inputs before classification~\cite{guo2018countering, Xu_2018}, as well as \emph{adversarial example detectors} that aim to flag suspicious inputs before or alongside the classifier~\cite{Hendrycks2017, KherchoucheFH22, Bunzel_2023_b, Bunzel_2023_d}. While adversarial training has become the most widely studied and benchmarked family of defenses, the standard evaluation paradigm for a new robust model is largely shared across the literature: it consists of reporting two numbers, the \emph{clean accuracy} (standard top-1 accuracy on unperturbed test data) and the \emph{robust accuracy} (accuracy under a standardized adversarial attack such as AutoAttack~\cite{croce2020autoattack}). These two numbers are then compared, either to each other or to prior robust models, to demonstrate progress.

What is almost always missing from this evaluation is the vanilla baseline. A reader who encounters a paper reporting, say, 93.6\% clean accuracy for an adversarially trained WideResNet~\cite{zagoruyko2016wideresnet} on CIFAR-10~\cite{krizhevsky2009cifar} has no immediate way of knowing what a \emph{non-adversarially-trained} model of the same architecture achieves, or, more broadly, what the state of the art in standard classification looks like. This omission matters because the entire value proposition of adversarial training research rests on a trade-off: how much standard accuracy does one sacrifice, and how much robustness does one gain? Without referencing the vanilla baseline, the cost side of this trade-off is invisible.

The problem is exacerbated by the fact that robustness papers typically use architectures (e.g., WideResNet-70-16~\cite{wang2023better}, WideResNet-34-10~\cite{SehwagMHD0CM22}) whose vanilla counterparts are not necessarily well-known or easily found. A fair comparison requires either (i)~matching the architecture exactly, (ii)~matching the temporal context (what was the best vanilla model at the time the robust model was published?), or (iii)~comparing against a broad reference distribution of vanilla models.

This information gap has direct business consequences. An organization deciding whether to adopt adversarially training must weigh the robustness benefit against the standard-accuracy cost: lower accuracy means more misclassifications, degraded user experience, potential revenue loss. Yet when a robustness paper reports a clean accuracy of, say, 84\% without referencing what a vanilla model achieves on the same task, the decision-maker cannot determine whether the 84\% represents a 2-point or a 10-point sacrifice. The difference between these two scenarios leads to qualitatively different business decisions: a 2-point gap may be an acceptable insurance premium against adversarial attacks, while a 10-point gap may render the model unusable for production. By making the gap explicit across the entire robustness literature, VanillaBench provides the missing quantitative basis for such decisions.

We address this gap with \textbf{VanillaBench}: a comprehensive evaluation that takes all 186~adversarially-trained models from RobustBench~\cite{croce2020robustbench} across four threat models and computes the clean-accuracy gap to multiple vanilla references. Specifically, our contributions are:

\begin{itemize}
  \item We assemble vanilla reference statistics (best, median, top-10 median) from Papers with Code~\cite{paperswithcode} for CIFAR-10~\cite{krizhevsky2009cifar}, CIFAR-100~\cite{krizhevsky2009cifar}, and ImageNet-1k~\cite{russakovsky2015imagenet}, computed over both all entries and no-extra-data entries.
  \item We construct a \emph{temporal} reference: the best vanilla accuracy achievable as of each robust model's publication year, enabling a year-matched comparison that controls for the rapid progress of standard classification.
  \item We curate an \emph{architecture-matched} reference: hand-matched standard accuracies for each robust model's underlying architecture, sourced from the torchvision model zoo~\cite{torchvision} for ImageNet, pytorch-cifar project~\cite{kuangliu_pytorchcifar} for CIFAR-10, pytorch-cifar100 project~\cite{weiaicunzai_pytorchcifar100} for CIFAR-100 and from the original architecture papers for CIFAR-10/100 and ImageNet.
  \item We compute $\Delta_{\mathrm{clean}}$, the gap between the robust model's clean accuracy and each vanilla reference, for all 186~models, and we show that this gap is universally negative and often large: even top-10 robust models lag the best vanilla model by 6--21~percentage points depending on the track.
\end{itemize}

\section{The VanillaBench Methodology}

\subsection{Robust Models}

We use the RobustBench leaderboard~\cite{croce2020robustbench} as our source of adversarially-trained models. RobustBench provides standardized, reliable robustness evaluations using AutoAttack~\cite{croce2020autoattack}, an ensemble of four attack methods that is widely accepted as a strong, parameter-free evaluation protocol. We consider all four tracks available at the time of this work:

\begin{itemize}
  \item \textbf{CIFAR-10~$\ell_\infty$~8/255:} 98~robust models, published 2018--2024.
  \item \textbf{CIFAR-10~$\ell_2$~0.5:} 20~robust models, published 2019--2024.
  \item \textbf{CIFAR-100~$\ell_\infty$~8/255:} 38~robust models, published 2019--2024.
  \item \textbf{ImageNet~$\ell_\infty$~4/255:} 30~robust models, published 2019--2024.
\end{itemize}

For each model, RobustBench records the clean accuracy (top-1 on the standard test set), the AutoAttack accuracy (top-1 under the standardized attack), the architecture, the publication venue and year, and whether the model was trained with additional data beyond the standard training set.

\subsection{Vanilla References}

We construct multiple vanilla reference points for each dataset, drawing from different sources.

\subsubsection{Papers with Code statistics.}
From the Papers with Code (PwC) leaderboards for image classification on CIFAR-10, CIFAR-100, and ImageNet-1k, we extract all entries and compute three statistics:
\begin{itemize}
  \item \textbf{Best}: the highest standard accuracy among the entries.
  \item \textbf{Median}: the median standard accuracy.
  \item \textbf{Top-10 median}: the median of the 10~highest-accuracy entries.
\end{itemize}
We compute each statistic over two subsets: (i)~\emph{all entries}, including models that leverage external pretraining on massive datasets such as JFT-300M~\cite{SunSSG17} or Instagram billion-scale data~\cite{MahajanGRHPLBM18}, and (ii)~\emph{no-extra-data entries} only, which exclude such models and thus provide a fairer comparison to robust models trained on the standard training set alone. This yields six reference points per track. The VanillaBench website additionally reports the mean, top-10 mean, and top-10 minimum for the all-entries subset.

Finally, we compute a \textbf{temporal best}: for each year~$y$, the best accuracy among \emph{all entries} published in or before~$y$, enabling a year-matched comparison that controls for the rapid progress of standard classification over time.

\subsubsection{Architecture-matched baselines.}
Because PwC statistics aggregate over heterogeneous architectures, they do not isolate the effect of adversarial training from architectural differences. To address this, we curate a hand-matched architecture table. For each robust model, we identify the underlying base architecture (e.g., WideResNet-70-16, ResNet-50~\cite{HeZRS16}, ConvNeXt-L~\cite{liu2022convnet}) and look up its standard accuracy from:
\begin{itemize}
  \item The torchvision model zoo (IMAGENET1K\_V1 weights~\cite{torchvision}) for ImageNet architectures, using the original published recipe (V1) rather than improved-recipe V2 weights.
  \item The pytorch-cifar~\cite{kuangliu_pytorchcifar} and pytorch-cifar100~\cite{weiaicunzai_pytorchcifar100} projects for CIFAR-10/100 data. 
  \item The original architecture papers for CIFAR and ImageNet data (e.g., Zagoruyko \& Komodakis~\cite{zagoruyko2016wideresnet} for WideResNet variants on CIFAR-10).
\end{itemize}
Where the robust model uses a modified architecture (e.g., RaWideResNet~\cite{PengXCHLDPMC23}, ConvStem-ViT~\cite{singh2023revisiting}), we match to the closest canonical base. When no exact vanilla number is published for an architecture, we use a \emph{proxy} match: a closely related architecture whose vanilla accuracy is known (e.g., WideResNet-70-16 is proxied to WideResNet-40-8~\cite{zagoruyko2016wideresnet}). Architecture matching (including proxy matches) is available for 184 of 186~robust models (98.9\%); of these, 90~(48.4\%) are direct matches and 94~are proxy matches. We report direct-match-only statistics separately to quantify the impact of proxy matching.

\subsection{The Accuracy Gap Metric}

For a robust model~$m$ with clean accuracy~$a_m$ and a vanilla reference value~$r$, we define the \emph{accuracy gap}:
\[
  \Delta_{\mathrm{clean}}(m, r) = a_m - r.
\]
A negative $\Delta_{\mathrm{clean}}$ indicates that the robust model sacrifices standard accuracy relative to the reference. We compute $\Delta_{\mathrm{clean}}$ for every robust model against every applicable reference. We emphasize that $\Delta_{\mathrm{clean}}$ captures only the \emph{cost} side of the robustness-accuracy trade-off. The \emph{benefit} side (robust accuracy) is reported in parallel but is not folded into a single scalar, since the trade-off is inherently two-dimensional.

\section{Results}

\subsection{Overview}

Table~\ref{tab:overview} presents the headline results: for each track, the number of robust models, the mean clean and robust accuracies, the best (all-entries) reference, and the mean $\Delta_{\mathrm{clean}}$ against five reference types. The gap is negative in every cell of the $\Delta$ columns. No robust model, on average, matches any vanilla reference.

\begin{table*}[t]
\centering
\caption{Overview of all four VanillaBench tracks. All references are computed over \emph{all entries} (including models that use additional training data). ``$\Delta$ best'' = mean $\Delta_{\mathrm{clean}}$ vs.\ best model; ``$\Delta$ median'' = vs.\ median; ``$\Delta$ t10med'' = vs.\ top-10 median; ``$\Delta$ arch'' = vs.\ architecture-matched baseline (including proxy matches); ``$\Delta$ year'' = vs.\ best model (all entries) as of the robust model's publication year. All values in percentage points.}
\label{tab:overview}
\small
\begin{tabular}{lrrrrrrrrr}
\toprule
Track & $N$ & Clean mean & Robust mean & Best (all) & $\Delta$ best & $\Delta$ median & $\Delta$ t10med & $\Delta$ arch & $\Delta$ year \\
\midrule
CIFAR-10 $\ell_\infty$ 8/255   & 98 & 87.49 & 53.44 & 99.50 & $-12.01$ & $-11.41$ & $-11.79$ & $-7.59$ & $-11.92$ \\
CIFAR-10 $\ell_2$ 0.5          & 20 & 91.79 & 76.36 & 99.50 & $-7.71$  & $-7.11$  & $-7.50$  & $-3.50$ & $-7.70$ \\
CIFAR-100 $\ell_\infty$ 8/255  & 38 & 66.61 & 32.12 & 96.08 & $-29.47$ & $-24.34$ & $-27.19$ & $-17.47$ & $-29.40$ \\
ImageNet $\ell_\infty$ 4/255   & 30 & 73.12 & 46.69 & 91.00 & $-17.88$ & $-9.48$  & $-16.73$ & $-8.52$ & $-17.76$ \\
\bottomrule
\end{tabular}
\end{table*}

Several patterns emerge:

\paragraph*{The best-vs.-median distinction matters.} The gap to the \emph{best} vanilla model is dramatically larger than the gap to the \emph{median} vanilla model, especially on ImageNet ($-17.88$ vs.\ $-9.48$). This reflects the wide spread of the PwC leaderboards, which range from 38.1\% (one shot or few shot) to 91.0\% on ImageNet. A robustness paper that reports clean accuracy without context can be compared favourably to the bottom of the distribution, making the gap appear smaller than it is.

\paragraph*{The temporal comparison is close to the overall best.} On all four tracks, the year-matched gap ($\Delta$~year) is nearly identical to the all-time best gap ($\Delta$~best). This is because the best vanilla model was published early (2019--2020 for CIFAR, 2022 for ImageNet) and remained the reference throughout. The median year-matched reference equals the all-time best on every track (99.50/99.50/96.08/91.00\%).

\paragraph*{The architecture-matched gap is smaller than the PwC gap.} On all four tracks, the architecture-matched gap is smaller than the best-vanilla gap, indicating that part of the apparent gap stems from comparing architecturally weaker robust models to stronger vanilla models. On CIFAR-10~$\ell_\infty$, the architecture-matched gap ($-7.59$) is about two-thirds of the best-vanilla gap ($-12.01$). On CIFAR-100 ($-17.47$), the architecture-matched gap is smaller than the median gap ($-24.34$), because the architecture-matched vanilla references (primarily WideResNet-28-10 at 80.75\% on CIFAR-100 and PreActResNet-18~\cite{he2016identity} at 72.92\%) are below the PwC median (90.95\%).

\paragraph*{The $\ell_2$ threat model has the smallest gap.} CIFAR-10~$\ell_2$~0.5 shows the smallest accuracy cost ($-7.71$ vs.\ best, $-3.50$ vs.\ architecture-matched), consistent with the fact that $\ell_2$-bounded perturbations at $\epsilon = 0.5$ are a weaker threat than $\ell_\infty$ at $8/255$.

\subsection{Per-Reference Detail}

Table~\ref{tab:detail} provides a more detailed breakdown, reporting mean and median $\Delta_{\mathrm{clean}}$ for each reference type on each track.

\begin{table*}[t]
\centering
\caption{Detailed $\Delta_{\mathrm{clean}}$ statistics (percentage points) for each reference type and track. ``$\Delta$ mean'' and ``$\Delta$ median'' refer to the distribution of $\Delta_{\mathrm{clean}}$ across all robust models in the track. ``All entries'' includes models that use additional training data; ``No extra data'' excludes them. ``$n$ (arch)'' = number of models with architecture-matched baselines available. ``Direct only'' = architecture-matched restricted to direct matches (excluding proxy-matched architectures). ``Year-matched best (all)'' = best model (all entries) as of the robust model's publication year. For the architecture-matched, Direct only, and Year-matched rows, the Ref.\ value is the median of per-model reference values.}
\label{tab:detail}
\small
\begin{tabular}{llrrrrr}
\toprule
Track & Reference & Ref.\ value & $\Delta$ mean & $\Delta$ median & $\Delta$ min & $\Delta$ max \\
\midrule
\multirow{9}{*}{CIFAR-10 $\ell_\infty$ 8/255}
  & Best (all entries)            & 99.50 & $-12.01$ & $-12.08$ & $-54.77$ & $-4.27$ \\
  & Median (all entries)          & 98.90 & $-11.41$ & $-11.48$ & $-54.17$ & $-3.67$ \\
  & Top-10 median (all entries)   & 99.28 & $-11.79$ & $-11.86$ & $-54.55$ & $-4.05$ \\
  & Best (no extra data)          & 99.50 & $-12.01$ & $-12.08$ & $-54.77$ & $-4.27$ \\
  & Median (no extra data)        & 98.70 & $-11.21$ & $-11.28$ & $-53.97$ & $-3.47$ \\
  & Top-10 median (no extra data) & 99.10 & $-11.61$ & $-11.67$ & $-54.37$ & $-3.87$ \\
  & Architecture-matched ($n\!=\!96$) & 96.00 & $-7.59$  & $-7.82$  & $-15.76$ & $-0.11$ \\
  & \quad Direct only ($n\!=\!41$)    & 96.00 & $-7.92$  & $-7.75$  & $-15.76$ & $-2.18$ \\
  & Year-matched best (all)       & 99.50 & $-11.92$ & $-12.00$ & $-54.77$ & $-4.27$ \\
\midrule
\multirow{9}{*}{CIFAR-10 $\ell_2$ 0.5}
  & Best (all entries)            & 99.50 & $-7.71$  & $-8.49$  & $-11.48$ & $-3.76$ \\
  & Median (all entries)          & 98.90 & $-7.11$  & $-7.90$  & $-10.88$ & $-3.16$ \\
  & Top-10 median (all entries)   & 99.28 & $-7.50$  & $-8.28$  & $-11.27$ & $-3.55$ \\
  & Best (no extra data)          & 99.50 & $-7.71$  & $-8.49$  & $-11.48$ & $-3.76$ \\
  & Median (no extra data)        & 98.70 & $-6.91$  & $-7.70$  & $-10.68$ & $-2.96$ \\
  & Top-10 median (no extra data) & 99.10 & $-7.31$  & $-8.09$  & $-11.08$ & $-3.36$ \\
  & Architecture-matched ($n\!=\!20$) & 95.34 & $-3.50$  & $-3.51$  & $-7.98$  & $+0.40$ \\
  & \quad Direct only ($n\!=\!9$)     & 95.11 & $-4.04$  & $-4.21$  & $-6.95$  & $-0.84$ \\
  & Year-matched best (all)       & 99.50 & $-7.70$  & $-8.48$  & $-11.48$ & $-3.76$ \\
\midrule
\multirow{9}{*}{CIFAR-100 $\ell_\infty$ 8/255}
  & Best (all entries)            & 96.08 & $-29.47$ & $-30.57$ & $-42.25$ & $-10.87$ \\
  & Median (all entries)          & 90.95 & $-24.34$ & $-25.45$ & $-37.12$ & $-5.74$ \\
  & Top-10 median (all entries)   & 93.80 & $-27.19$ & $-28.30$ & $-39.97$ & $-8.59$ \\
  & Best (no extra data)          & 96.08 & $-29.47$ & $-30.57$ & $-42.25$ & $-10.87$ \\
  & Median (no extra data)        & 90.00 & $-23.39$ & $-24.49$ & $-36.17$ & $-4.79$ \\
  & Top-10 median (no extra data) & 92.76 & $-26.15$ & $-27.25$ & $-38.92$ & $-7.55$ \\
  & Architecture-matched ($n\!=\!38$) & 80.75 & $-17.47$ & $-16.89$ & $-34.48$ & $-6.90$ \\
  & \quad Direct only ($n\!=\!10$)    & 78.18 & $-13.96$ & $-13.74$ & $-21.52$ & $-6.90$ \\
  & Year-matched best (all)       & 96.08 & $-29.40$ & $-30.57$ & $-42.25$ & $-10.87$ \\
\midrule
\multirow{9}{*}{ImageNet $\ell_\infty$ 4/255}
  & Best (all entries)            & 91.00 & $-17.88$ & $-15.72$ & $-38.08$ & $-9.52$ \\
  & Median (all entries)          & 82.60 & $-9.48$  & $-7.32$  & $-29.68$ & $-1.12$ \\
  & Top-10 median (all entries)   & 89.85 & $-16.73$ & $-14.57$ & $-36.93$ & $-8.37$ \\
  & Best (no extra data)          & 91.00 & $-17.88$ & $-15.72$ & $-38.08$ & $-9.52$ \\
  & Median (no extra data)        & 82.50 & $-9.38$  & $-7.22$  & $-29.58$ & $-1.02$ \\
  & Top-10 median (no extra data) & 89.85 & $-16.73$ & $-14.57$ & $-36.93$ & $-8.37$ \\
  & Architecture-matched ($n\!=\!30$) & 83.12 & $-8.52$  & $-7.67$  & $-20.51$ & $+0.39$ \\
  & \quad Direct only ($n\!=\!30$)    & 83.12 & $-8.52$  & $-7.67$  & $-20.51$ & $+0.39$ \\
  & Year-matched best (all)       & 91.00 & $-17.76$ & $-16.34$ & $-37.28$ & $-9.52$ \\
\bottomrule
\end{tabular}
\end{table*}

Several observations stand out:

\begin{itemize}
  \item On CIFAR-10~$\ell_\infty$, the minimum $\Delta_{\mathrm{clean}}$ against the best vanilla model is $-54.77$~pp, indicating that some robust models sacrifice more standard accuracy than they retain in robust accuracy. The maximum is $-4.27$~pp, achieved by Bai et al.~\cite{bai2024mixednuts} (MixedNUTS), which uses a hybrid approach combining a robust and a non-robust classifier.
  \item On ImageNet, no robust model exceeds any aggregate vanilla reference. The smallest gap against any PwC reference is $-1.02$~pp (vs.\ no-extra-data median), achieved by Singh et al.~\cite{singh2023revisiting} (ConvNeXt-B + ConvStem).
  \item The architecture-matched comparison reveals only two cases where a robust model \emph{exceeds} its matched vanilla baseline: $+0.40$ on CIFAR-10~$\ell_2$ and $+0.39$ on ImageNet~$\ell_\infty$. Both are small positive margins. The earlier $+21.08$ outlier (Sehwag et al.~\cite{sehwag2022proxy}, ResNet-18) has been resolved by updating the ResNet-18 CIFAR-10 baseline to its correct value (93.02\%).
  \item The direct-match-only statistics (``Direct only'' rows) exclude proxy-matched architectures. On CIFAR-10~$\ell_\infty$ and $\ell_2$, the direct-match gaps are similar to or slightly larger than the full gaps, suggesting that proxy matching does not systematically bias the results on these tracks.
\end{itemize}

\subsection{Top-10 Robust Models}

Table~\ref{tab:top10} focuses on the 10~most robust models (by AutoAttack accuracy) per track, which represent the state of the art in adversarial training. Even among these best-performing models, the accuracy gap remains substantial.

\begin{table*}[t]
\centering
\caption{Top-10 robust models per track: mean and median $\Delta_{\mathrm{clean}}$ (pp) against each no-extra-data vanilla reference.}
\label{tab:top10}
\small
\begin{tabular}{lrrrrrrrr}
\toprule
Track & Top-10 clean mean & \multicolumn{2}{c}{$\Delta$ vs.\ best} & \multicolumn{2}{c}{$\Delta$ vs.\ median} & \multicolumn{2}{c}{$\Delta$ vs.\ top-10 median} \\
& & mean & median & mean & median & mean & median \\
\midrule
CIFAR-10 $\ell_\infty$ 8/255  & 93.65 & $-5.85$ & $-6.07$ & $-5.05$ & $-5.27$ & $-5.45$ & $-5.66$ \\
CIFAR-10 $\ell_2$ 0.5         & 93.80 & $-5.70$ & $-5.15$ & $-4.90$ & $-4.35$ & $-5.30$ & $-4.75$ \\
CIFAR-100 $\ell_\infty$ 8/255 & 74.87 & $-21.21$ & $-21.59$ & $-15.13$ & $-15.51$ & $-17.88$ & $-18.27$ \\
ImageNet $\ell_\infty$ 4/255  & 77.94 & $-13.06$ & $-13.03$ & $-4.56$ & $-4.53$ & $-11.91$ & $-11.88$ \\
\bottomrule
\end{tabular}
\end{table*}

Even the best-available robust models sacrifice 4.6--15.1~pp of standard accuracy relative to the median vanilla model. The gap widens to 5.7--21.2~pp relative to the best vanilla model. This confirms that the robustness--accuracy trade-off is not merely a problem for weaker models; it affects the entire frontier.

\subsection{Temporal Trends}

\begin{figure*}[t]
\centering
\begin{tabular}{ccc}
\includegraphics[width=0.31\textwidth]{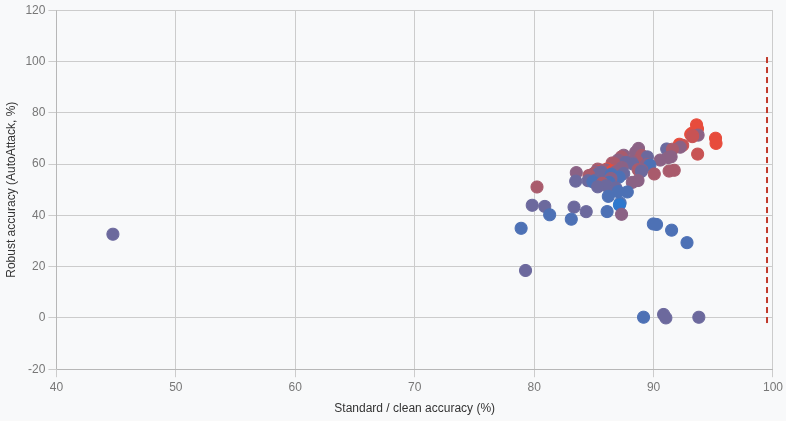} &
\includegraphics[width=0.31\textwidth]{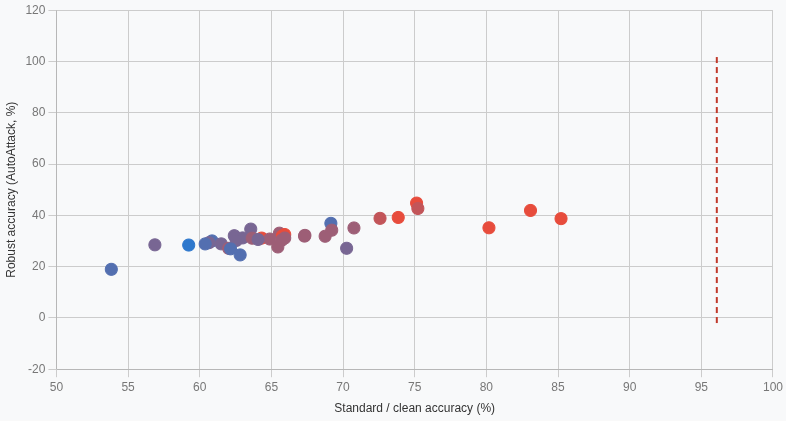} &
\includegraphics[width=0.31\textwidth]{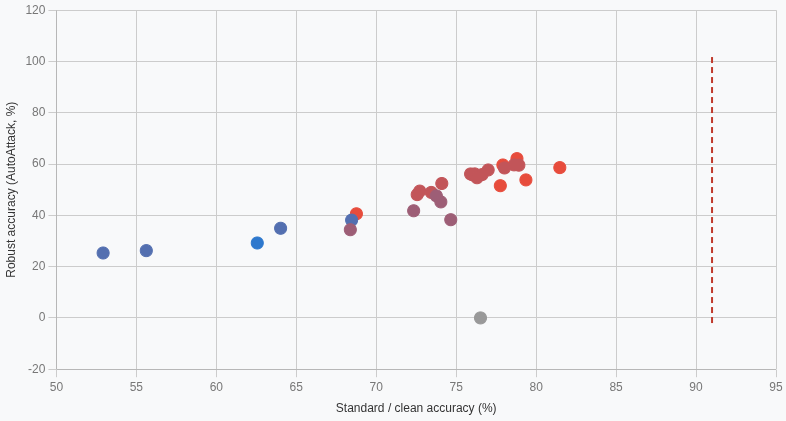} \\
(a) CIFAR-10 $\ell_\infty$ 8/255 & (b) CIFAR-100 $\ell_\infty$ 8/255 & (c) ImageNet $\ell_\infty$ 4/255 \\
\end{tabular}
\caption{Clean accuracy vs.\ robust accuracy (AutoAttack) scatter plots for all $\ell_\infty$ robust models. Each point represents one model; colour indicates publication year (blue = older $\approx$~2017--2019, red = newer $\approx$~2023--2024). (a) CIFAR-10 $\ell_\infty$ ($\epsilon=8$): no robust model exceeds 64\% clean accuracy; vanilla top-10 median reference at 99.44\%. (b) CIFAR-100 $\ell_\infty$ ($\epsilon=8$): similar gap, with robust models between 16--54\% robust accuracy. (c) ImageNet $\ell_\infty$ ($\epsilon=4$): robust models cluster between 26--64\% robust accuracy, far below the vanilla reference at 93.58\%. The vertical gap to the vanilla line is the $\Delta_{\mathrm{clean}}$ metric central to VanillaBench.}
\label{fig:scatter}
\end{figure*}

\begin{figure*}[t]
\centering
\begin{tabular}{cc}
\includegraphics[width=0.45\textwidth]{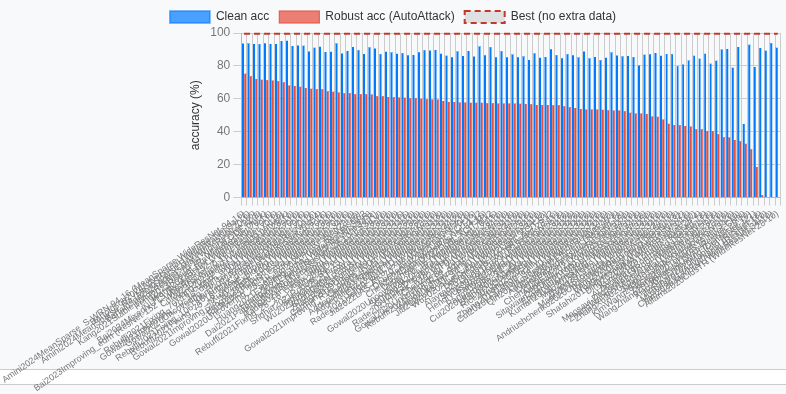} &
\includegraphics[width=0.45\textwidth]{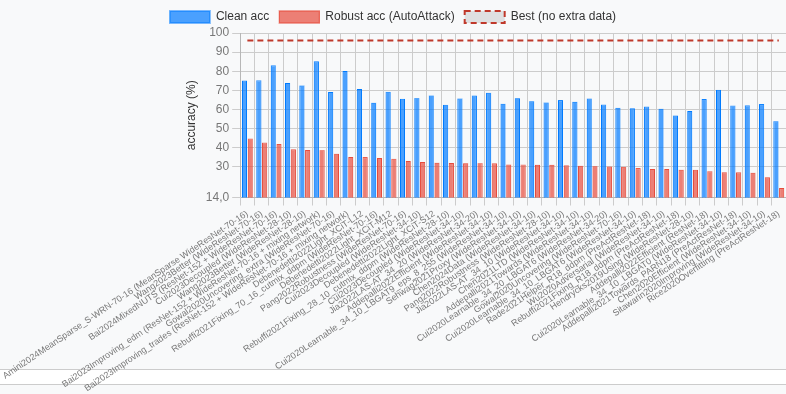} \\
(a) CIFAR-10 $\ell_\infty$ 8/255 & (b) CIFAR-100 $\ell_\infty$ 8/255 \\
\includegraphics[width=0.45\textwidth]{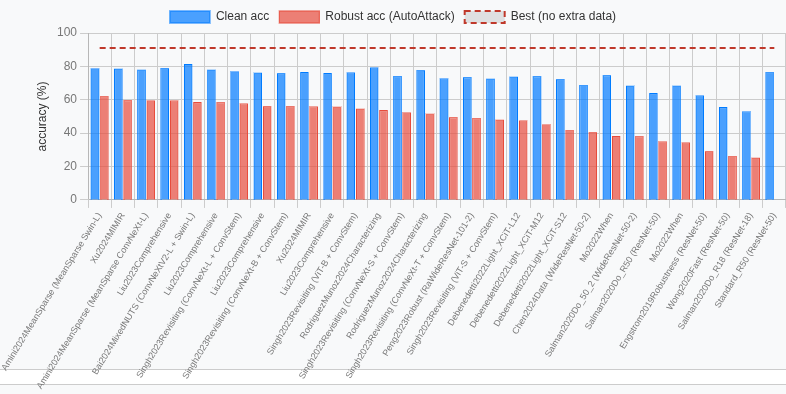} &
\includegraphics[width=0.45\textwidth]{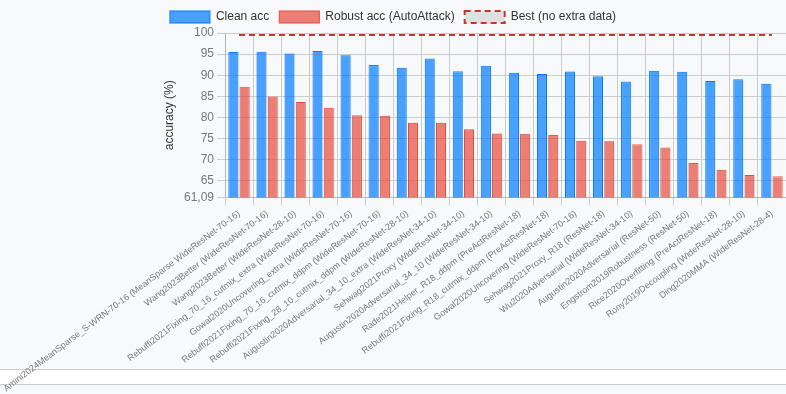} \\
(c) ImageNet $\ell_\infty$ 4/255 & (d) CIFAR-10 $\ell_2$ 0.5 \\
\end{tabular}
\caption{Clean accuracy (blue) and AutoAttack robust accuracy (red) bar charts for robust models per track. (a) CIFAR-10 $\ell_\infty$ ($\epsilon=8$): robust models achieve 27--64\% robust accuracy, while clean accuracy ranges from 82--94\%. The red dashed line marks 99.44\% (top-10 median vanilla reference). (b) CIFAR-100 $\ell_\infty$ ($\epsilon=8$): robust accuracy 16--54\%, clean accuracy 79--95\%, reference at 92.12\%. (c) ImageNet $\ell_\infty$ ($\epsilon=4$): robust accuracy 26--64\%, clean accuracy 50--84\%, reference at 93.58\%. (d) CIFAR-10 $\ell_2$ ($\sigma=0.5$): the only track where some robust models achieve comparable or superior clean accuracy relative to the vanilla reference (99.44\%). The vertical gap between the blue bar and the dashed line is the $\Delta_{\mathrm{clean}}$ that VanillaBench makes explicit, quantifying the clean accuracy sacrifice imposed by adversarial training.}
\label{fig:bar}
\end{figure*}

\begin{figure*}[t]
\centering
\begin{tabular}{cc}
\includegraphics[width=0.45\textwidth]{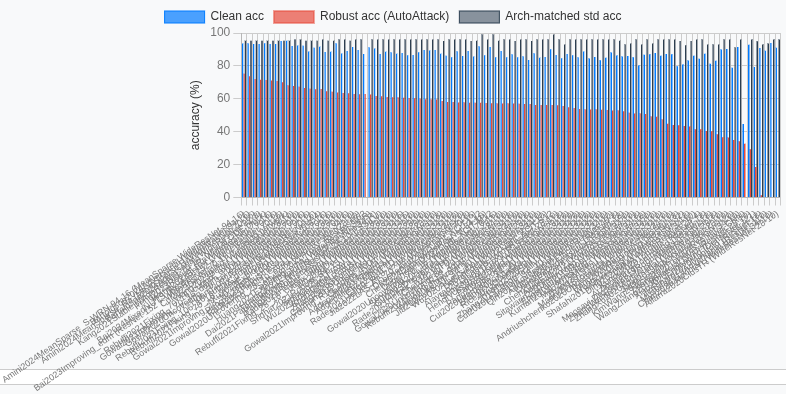} &
\includegraphics[width=0.45\textwidth]{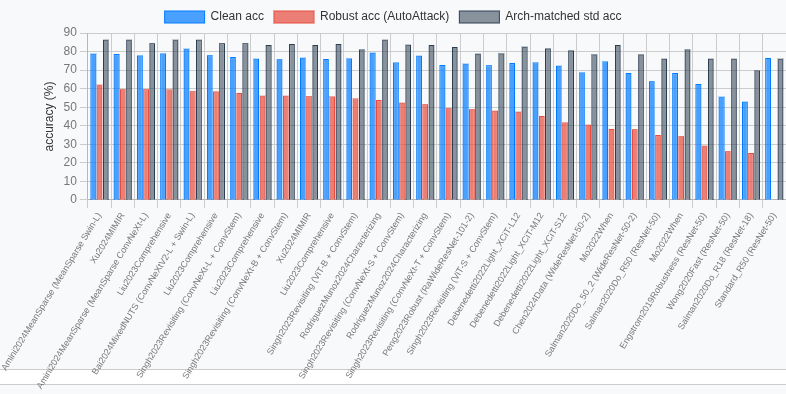} \\
(a) CIFAR-10 $\ell_\infty$ 8/255 & (b) ImageNet $\ell_\infty$ 4/255 \\
\end{tabular}
\caption{Architecture-matched comparison: per-model bar charts showing clean accuracy (blue), AutoAttack robust accuracy (red), and architecture-matched vanilla accuracy (dark grey). The grey bar represents a vanilla model sharing the same architecture (or proxy architecture) as the robust model, isolating the clean accuracy cost of adversarial training from architectural differences. (a) CIFAR-10 $\ell_\infty$ ($\epsilon=8$): no robust model exceeds its architecture-matched vanilla baseline; gaps range from 4.8--7.8~pp for directly matched models. (b) ImageNet $\ell_\infty$ ($\epsilon=4$): larger gaps, with robust models falling 8.5--20~pp below their matched vanilla counterpart (direct match only). These results confirm that even when controlling for architecture, adversarial training incurs a substantial clean accuracy penalty. Proxy-matched models (94 total) use the nearest available vanilla architecture when an exact match is unavailable; direct-match-only analysis is reported in Table~2.}
\label{fig:arch}
\end{figure*}

Table~\ref{tab:temporal} breaks down the mean clean accuracy of robust models by publication year, alongside the year-matched best (all entries).

\begin{table}[t]
\centering
\caption{Mean clean accuracy of robust models by publication year, compared to the year-matched best (all entries) accuracy.}
\label{tab:temporal}
\small
\begin{tabular}{lrrrr}
\toprule
Year & \multicolumn{2}{c}{CIFAR-10 $\ell_\infty$} & \multicolumn{2}{c}{ImageNet $\ell_\infty$} \\
 & Robust mean & Vanilla best & Robust mean & Vanilla best \\
\midrule
2018 & 87.14 & 93.57 & --- & 80.62 \\
2019 & 86.99 & 99.37 & 62.56 & 87.54 \\
2020 & 84.56 & 99.50 & 60.25 & 90.20 \\
2021 & 88.39 & 99.50 & --- & 90.20 \\
2022 & 87.50 & 99.50 & 72.64 & 91.00 \\
2023 & 93.16 & 99.50 & 75.88 & 91.00 \\
2024 & 91.42 & 99.50 & 77.35 & 91.00 \\
\bottomrule
\end{tabular}
\end{table}

On CIFAR-10, the vanilla best was already at 99.37\% in 2019 (BiT-L~\cite{kolesnikov2019bit}), while the best mean robust clean accuracy in any year is 93.16\% (2023). On ImageNet, the vanilla best reached 91.0\% in 2022 (CoCa~\cite{yu2022coca}), while the best mean robust clean accuracy is 77.35\% (2024). The gap has narrowed over time but remains at 6+~pp on CIFAR-10 and 13+~pp on ImageNet even in the most recent year.

\subsection{Per-Track Analysis}

\subsubsection{CIFAR-10 $\ell_\infty$ 8/255.}
This is the most populated track with 98~models. The mean $\Delta_{\mathrm{clean}}$ against the best vanilla model is $-12.01$~pp, and even the mean gap of the top-10 most robust models is $-5.85$~pp. The architecture-matched gap is smaller ($-7.59$~pp, median reference 96.00\%), because many robust models use WideResNet-70-16~\cite{wang2023better, PengXCHLDPMC23, bai2024mixednuts} (vanilla: 95.34\%, proxy-matched to WideResNet-40-8) or WideResNet-28-10~\cite{wang2023better, XuSGGH23, CuiTZ00Z24} (vanilla: 96.0\%), architectures whose standard accuracy is high but not at the 99.5\% SOTA level. The most robust model, Amini et al.~\cite{amini2024meansparse} (MeanSparse WideResNet-94-16), achieves 93.6\% clean and 75.28\% robust accuracy. This is $-5.90$~pp below the best vanilla model.

\subsubsection{CIFAR-10 $\ell_2$ 0.5.}
This track has the smallest accuracy cost. The mean $\Delta_{\mathrm{clean}}$ against the best vanilla model is $-7.71$~pp, and the architecture-matched gap is $-3.50$~pp (median reference 95.34\%). The top model, Amini et al.~\cite{amini2024meansparse} (MeanSparse WideResNet-70-16), achieves 95.51\% clean and 87.28\% robust accuracy. This is $+0.17$~pp above its proxy-matched vanilla baseline (WideResNet-40-8, 95.34\%). This is consistent with the weaker perturbation budget of $\ell_2 = 0.5$.

\subsubsection{CIFAR-100 $\ell_\infty$ 8/255.}
This track exhibits the largest accuracy gap: $-29.47$~pp mean against the best vanilla model. The architecture-matched gap is $-17.47$~pp, smaller than the PwC median gap ($-24.34$~pp), because the architecture-matched vanilla references (primarily WideResNet-28-10 at 80.75\% (the median architecture-matched reference) on CIFAR-100 and PreActResNet-18 at 72.92\%) are below the PwC median of 90.95\%. The top model, Amini et al.~\cite{amini2024meansparse} (MeanSparse WideResNet-70-16), achieves 75.13\% clean and 44.78\% robust accuracy. This is $-20.21$~pp below its proxy-matched vanilla baseline. The CIFAR-100 results underscore that the robustness-accuracy trade-off is dataset-dependent and that progress on the harder 100-class setting has been much slower.

\subsubsection{ImageNet $\ell_\infty$ 4/255.}
On ImageNet, the mean $\Delta_{\mathrm{clean}}$ is $-17.88$~pp against the best vanilla model but only $-9.48$~pp against the median vanilla model (all entries). This wide spread reflects the diversity of the ImageNet leaderboard (105~vanilla entries ranging from 38.1\% to 91.0\%). The architecture-matched gap ($-8.52$~pp, median reference 83.12\%) is close to the PwC median gap ($-9.48$~pp), suggesting that on ImageNet, architectural matching does not substantially change the picture. The top model, Amini et al.~\cite{amini2024meansparse} (MeanSparse Swin-L), achieves 78.80\% clean and 62.12\% robust accuracy. This is $-12.20$~pp below the best vanilla model. Notably, 29 of 30~ImageNet robust models use no additional training data, making this the ``cleanest'' track in terms of training setup.

\section{Discussion}

\subsection{Why the Vanilla Gap Is Rarely Shown}

The absence of vanilla baselines in robustness papers is not malicious. It reflects several practical and structural factors. First, adversarial training papers are written for the robustness community, where the implicit baseline is prior robust models, not vanilla models. Second, the architectures used for adversarial training (e.g., WideResNet-70-16) are often not the same as those that dominate standard leaderboards (e.g., ViT~\cite{dosovitskiy2020vit}, ConvNeXt), making direct comparison awkward. Third, many robust models use additional data (26.5\% on CIFAR-10~$\ell_\infty$, 23.7\% on CIFAR-100), which further complicates the comparison. Should the vanilla reference also use additional data?

VanillaBench addresses these complications by providing multiple references that span different notions of ``vanilla.'' The best and top-10 median references show the gap to the absolute state of the art. The median reference shows the gap to the ``typical'' model. Each of these is computed over both all entries and no-extra-data entries, so the reader can assess whether the use of additional training data materially changes the picture. The year-matched reference controls for temporal progress. The architecture-matched reference isolates the adversarial training effect. Together, these provide a much richer picture than any single comparison.

\subsection{The Two-Dimensional Trade-Off}

We deliberately do not collapse $\Delta_{\mathrm{clean}}$ and robust accuracy into a single scalar (e.g., a weighted average). The robustness--accuracy trade-off is fundamentally two-dimensional, and any scalarization hides information. Figure~\ref{fig:scatter} makes the trade-off visible: each point occupies a position in (clean, robust) space, and the vanilla reference line provides the x-axis anchor that is typically missing. Researchers and practitioners can use this visualization to assess whether a given model's trade-off point is ``good'' relative to both axes simultaneously.

\subsection{Threats to Validity}

\textbf{Architecture matching imperfections.} Our architecture-matched baselines are hand-curated and sometimes based on values from the original architecture papers rather than standardized re-evaluations. Additionally, 94 of 184~architecture-matched entries are \emph{proxy matches}. The robust model's exact architecture has no published vanilla number, so a closely related architecture is used instead (e.g., WideResNet-70-16 is proxied to WideResNet-40-8). We report direct-match-only statistics separately to quantify the impact of this proxy matching.

\textbf{Papers with Code curation.} The PwC leaderboards have been curated/trimmed by Hugging Face, so the reference statistics reflect the live curated boards rather than an exhaustive historical list. The number of vanilla entries is relatively small for CIFAR (20--22), which means the statistics are sensitive to the inclusion or exclusion of individual entries.

\textbf{Additional data flag.} Our ``no extra data'' filter relies on PwC's \texttt{uses\_additional\_data} flag, which may not perfectly capture all nuances (e.g., self-supervised pretraining on the same dataset is sometimes considered ``additional data'' and sometimes not).

\textbf{AutoAttack as the sole robustness metric.} We use the AutoAttack accuracy from RobustBench as the robustness measure. AutoAttack is a strong standardized attack, but robust accuracy under other threat models (e.g., corrupted inputs, distribution shift) may tell a different story.

\section{Related Work}

\textbf{Adversarial robustness benchmarks.} RobustBench~\cite{croce2020robustbench} is the most widely used standardized benchmark for adversarial robustness, providing a curated leaderboard of models evaluated under AutoAttack. However, RobustBench focuses exclusively on robust models and does not provide vanilla baselines. VanillaBench complements RobustBench by adding the missing vanilla reference layer.

\textbf{The accuracy--robustness trade-off.} The fundamental tension between standard and robust accuracy was noted early on~\cite{tsipras2018robustness,zhang2019theoretically}. Tsipras et al.~\cite{tsipras2018robustness} provided a theoretical analysis showing that the trade-off is inherent, not merely an artifact of suboptimal training. Zhang et al.~\cite{zhang2019theoretically} decomposed robust and standard error into boundary and interior components. However, these works analyze the trade-off theoretically or on small-scale experiments. They do not systematically quantify the gap across the entire robustness literature against vanilla references.

\textbf{Robustness without accuracy loss.} Several works have argued that the trade-off can be ``reconciled''~\cite{pang2022robustness} or mitigated through better training procedures~\cite{wang2023better,bai2024mixednuts}. Our results confirm that progress has been made. The most recent models are closer to vanilla accuracy than early ones, but the gap is still substantial, especially on harder datasets like CIFAR-100.

\textbf{Standard classification benchmarks.} Papers with Code~\cite{paperswithcode} provides the standard accuracy leaderboards that we use as vanilla references. The torchvision model zoo~\cite{torchvision} provides verified, reproducible standard accuracies for common architectures.

\section{Conclusion \& Future Work}

We have presented VanillaBench, a systematic evaluation of the accuracy cost of adversarial robustness. By computing the clean-accuracy gap between every RobustBench model and multiple vanilla references, including best, median, top-10 median (all entries and no-extra-data), year-matched, and architecture-matched, we make explicit a trade-off that is typically only implicit in the literature. The gap is universally negative. Across 186~models and four threat models, no robust model matches the best vanilla model, and even top-10 robust models sacrifice 4.6--15.1~pp of standard accuracy relative to the median vanilla model. The architecture-matched comparison, which isolates the adversarial training effect, confirms that the gap is not merely an artifact of comparing different architectures. We recommend that future robustness papers report vanilla-referenced accuracy gaps as a standard component of evaluation, and we provide the VanillaBench tooling\footnote{\url{https://bunni90.github.io/robust-vs-vanilla.html}} to facilitate this.

Several directions remain for future work. First, the vanilla reference accuracies used in this work are sourced from published papers and model zoos, which may reflect different evaluation protocols, preprocessing pipelines, or training recipes. Conducting standardized in-house evaluations of both robust and vanilla models under a unified protocol would eliminate this source of variability and yield more directly comparable gap estimates. Second, a number of architectures that appear frequently in adversarial training (e.g., WideResNet-70-16, WideResNet-94-16) lack published vanilla accuracies, forcing us to rely on proxy matches. Training and evaluating vanilla versions of these architectures would remove the need for proxy matching and tighten the architecture-matched comparison. Third, the most recent model in RobustBench dates to 2024, and the benchmark datasets (CIFAR-10, CIFAR-100, ImageNet-1k) have remained unchanged for years. Extending VanillaBench to encompass newer robust models and additional, more modern evaluation datasets would keep the benchmark current and broaden its applicability.

\section*{Acknowledgments}

This work was supported by the ATHENE flagship project funded by the German Federal Ministry of Education and Research (BMBF).

\bibliographystyle{ACM-Reference-Format}
\bibliography{refs}

\end{document}